\documentclass[ aip, jmp, amsmath,amssymb,
 reprint,%
]{revtex4-1}

\usepackage{graphicx}
\usepackage{dcolumn}
\usepackage{bm}

\begin{document}
\title{The influence of  anisotropic gate potentials on the phonon induced spin-flip rate   in GaAs  quantum dots }

\author{Sanjay Prabhakar}
\affiliation{M\,$^2$NeT Laboratory,Wilfrid Laurier University, Waterloo, Ontario N2L 3C5, Canada}
\author{Roderick V. N. Melnik}
\affiliation{M\,$^2$NeT Laboratory,Wilfrid Laurier University, Waterloo, Ontario N2L 3C5, Canada}
 \affiliation{%
Gregorio Millan Institute, Universidad Carlos III de Madrid, 28911, Leganes, Spain
}%
\author{Luis L. Bonilla$^{2}$}

\date{December 16, 2011}
\begin{abstract}
We study the anisotropic orbital effect in the  electric field tunability of the phonon induced  spin-flip rate in quantum dots (QDs). Our study shows that anisotropic gate potential enhances the spin-flip rate and reduces the level crossing point to a lower QDs radius due to the suppression of the Land$\acute{e}$ g-factor towards bulk crystal. In the range of $10^4-10^6$ V/cm, the electric field tunability of the phonon induced spin-flip rate can be manipulated through strong Dresselhaus spin-orbit coupling. These results might assist the development of a  spin based  solid state quantum computer by manipulating  spin-flip rate through spin-orbit coupling  in a regime where the g-factor changes its sign.
\end{abstract}

\maketitle

Controlling the single electron spins in QDs  through the application of anisotropic gate potential is important for the design of solid state quantum computer.~\cite{takahashi10,kanai11,marquardt11} Tunability of the  phonon induced spin-flip rate and the electron g-factor in III-V semiconductor QDs can be manipulated through the application of externally applied gate potentials.~\cite{sousa03,khaetskii01,prabhakar09,pryor06} The strength of the Rashba-Dresselhaus spin-orbit coupling is determined by the asymmetric triangular quantum well along z-direction which approximately estimates the density of electrons at the heterojunction.~\cite{sousa03,prabhakar10} Rashba spin-orbit coupling arises from the structural inversion asymmetry along the growth direction and Dresselhaus spin-orbit coupling arises from bulk inversion asymmetry in the crystal lattice.~\cite{bychkov84,dresselhaus55}

Recently, it has been made possible to measure the electron spin states in gated QDs in presence of magnetic fields along arbitrary direction.~\cite{takahashi10} The physics behind this has been theoretically  investigated by authors in Refs.~\onlinecite{nowak11,prabhakar11} which confirms that the spin-orbit coupling can be used as a control parameter in the electric field and magnetic field tunability of the electron g-factor tensor.
The authors in Refs.~\onlinecite{elzerman04,kroutvar04}
measured the long spin relaxation times, approximately  0.85 ms in GaAs QDs by pulsed relaxation rate measurements and approximately 20 ms in InGaAs QDs by  optical orientation measurements. These spin-flip rate measurements in QDs confirm the theoretical predictions of the suppression of the phonon induced spin-flip rate by spin-orbit coupling with respect to the environment.~\cite{golovach04,khaetskii00,prabhakar10} Our work is along the lines of   Refs.~\onlinecite{sousa03,bulaev05,bulaev05a} with several new important findings. In particular, in  this paper, we study  the anisotropic orbital effect on the phonon induced spin-flip rate  for a system where the area of the symmetric and asymmetric QDs kept constant. Based on both theoretical and finite element numerical simulation methods, we find that the anisotropic potential enhances the spin flip rate and reduces the level crossing point to a lower quantum dot radius due to the suppression of the g-factor towards bulk crystal.
\begin{figure}
\includegraphics[width=9cm,height=7cm]{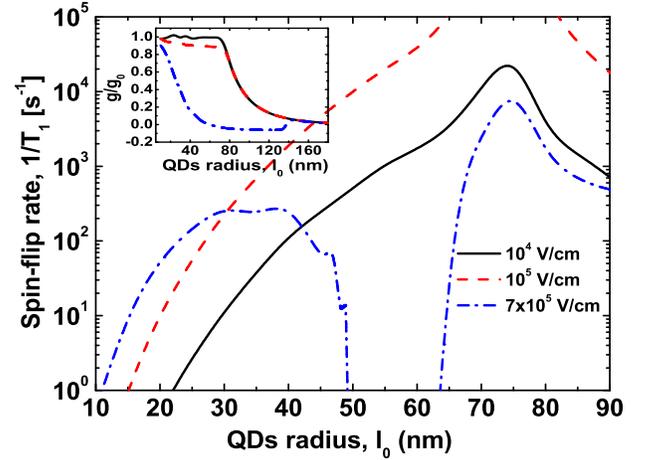}
\caption{\label{fig1}
(Color online) Phonon induced spin-flip rate due to spin-orbit admixture mechanism as a function of QDs radius in symmetric QDs (a=b=1). We choose B=1 Tesla. Also, inset plot  shows the g-factor vs. QDs radius. The level crossing point occurs at $\ell_0=73$ nm. The material constants for GaAs QDs have been chosen from Refs.~\onlinecite{sousa03,cardona88} as follows:
$g_0=-0.44$, $m=0.067$, $\gamma_R=4.4 ~\mathrm{{\AA}^2}$, $\gamma_D=26~\mathrm{eV{\AA}^3}$, $eh_{14}=2.34\times 10^{-5}~\mathrm{erg/cm}$, $s_l=5.14\times 10^{5}~\mathrm{cm/s}$, $s_t=3.03\times 10^{5}~\mathrm{cm/s}$ and $\rho=5.3176 ~\mathrm{g/cm^3}$. At $E=7\times 10^5$ V/cm shown by dashed-dotted lines, the admixture mechanism due to spin-orbit coupling on the spin-flip rate  is quite different because the electron spin states  change their sign in these regime (see inset plot).
}
\end{figure}

We consider 2D anisotropic  semiconductor QDs formed in the conduction band in the presence of magnetic field  B along z-direction. The total Hamiltonian $H = H_{xy} +  H_{so}$ of an electron in the conduction band  under the Kane model~\cite{pikus95,khaetskii01} can be written as
\begin{eqnarray}
H_{xy} = {\frac {\vec{P}^2}{2m}} + {\frac{1}{2}} m \omega_o^2
(a x^2 + b y^2) + {\frac \hbar 2}  \sigma_z \omega_z,\label{hxy}\\
H_{so} =\frac{\alpha_R}{\hbar}\left(\sigma_x P_y - \sigma_y P_x\right)+ \frac{\alpha_D}{\hbar}\left(-\sigma_x P_x + \sigma_y P_y\right),\label{rashba-dresselhaus}\\
\alpha_R=\gamma_ReE,~~~~~\alpha_D=0.78\gamma_D\left(\frac{2me}{\hbar^2}\right)^{2/3}E^{2/3},\label{coefficient-R-D}
\end{eqnarray}
where $\vec{P} = -i\hbar\mathbf{\nabla} + e \vec{A}$ is  the 2D electron momentum operator in the asymmetric gauge $\vec{A} = {\frac {B}{\sqrt a+\sqrt b}} (-y\sqrt b,x\sqrt a,0)$ and $\omega_z=g_0\mu_B B/\hbar$ is the Zeeman frequency. Also, $m$ is the effective mass,  $\mu_B$ is the Bohr magneton, $\vec{\sigma}$ is the Pauli spin matrices, $\omega_0=\frac{\hbar}{m\ell_0^2}$ is the strength of the parabolic confining potential with  quantum dot radius $\ell_0$.
The externally applied gate potential ($V_{xy}=1/2m\omega_o^2(ax^2+by^2)$) in our theoretical model defines the lateral
size of the QDs along x- and y-directions in the plane of 2DEG. By chosing $\ell_0=10-100$ nm, we mimic the  experimentally reported lateral
size of the QDs in Refs.~\onlinecite{takahashi10,thornton98} and the potential induced in this range is much lower than the break down
voltage ($400$ kV/cm) in GaAs heterojunctions.
The strength of the Rashba and Dresselhaus spin-orbit couplings is determined by the relation $\alpha_R/\alpha_D=1.5\times 10^{-3}E^{1/3}$ which tells us that $\alpha_R=\alpha_D$ at the electric field $E=3\times 10^6$ V/cm. In the range of $10^4-10^6$ V/cm, only the  Dresselhaus spin-orbit coupling has an appreciable contribution in the manipulation of spin-flip rate in QDs.
The asymmetric triangular quantum well potential ($E=-dV/dz$) arises  along the growth direction and
usually has a major contribution to the E-filed tunability of the g-factor and spin-flip rate due to the interplay
between Rashba-Dresselhaus spin-orbit couplings. The expression $(2meE/\hbar^2)^{-1/3}$ estimates the average
thickness of the 2DEG
where one can estimate the vertical average height of the QDs.~\cite{sousa03,prabhakar11}
By chosing  $E=(0.1-10)10^5$ V/cm, we estimate the average height of the QDs from $2$ nm to $10$ nm which is in the range of experimentally reported
values.~\cite{thornton98}

The above Hamiltonian~(\ref{hxy}) can be exactly diagonalized~\cite{schuh85,galkin04} and spin-orbit Hamiltonian can be used perturbatively to find the energy states of the QDs. The  energy spectrum can be written as
\begin{eqnarray}
H_{xy}=\left(n_++n_-+1\right)\hbar\omega_++\left(n_+-n_-\right)\hbar\omega_-+ {\frac \hbar 2}  \sigma_z \omega_z, \label{epsilon-a}\nonumber\\
H_{so}=\alpha_R \left(1+i\right)[b^{1/4}\kappa_+\left(s_+-i\right)a_++b^{1/4}\kappa_+\left(s_-+i\right)a_-\nonumber\\
+a^{1/4}\eta_-\left(i-s_-\right)a_++a^{1/4}\eta_-\left(i+s_+\right)a_-]\nonumber\\
+\alpha_D
\left(1+i\right)[a^{1/4}\kappa_-\left(i-s_-\right)a_++a^{1/4}\kappa_-\left(i+s_+\right)a_-\nonumber\\
+b^{1/4}\eta_+\left(-i+s_+\right)a_++b^{1/4}\eta_+\left(i+s_-\right)a_-]+H.c.,~~~~~~
\label{H-R}\nonumber\\
\kappa_{\pm}=\frac{1}{2\left(s_+-s_-\right)}\left\{\frac{1}{\ell}\sigma_x\pm i\frac{eB\ell}{\hbar}\left(\frac{1}{\sqrt{a}+\sqrt{b}}\right)\sigma_y\right\},\nonumber\\
\eta_{\pm}=\frac{1}{2\left(s_+-s_-\right)}\left\{\frac{1}{\ell}\sigma_y\pm i\frac{eB\ell}{\hbar}\left(\frac{1}{\sqrt{a}+\sqrt{b}}\right)\sigma_x\right\},\nonumber
\end{eqnarray}
and H.c. represents the Hermitian conjugate.  Also, $\omega_{\pm}=\frac{1}{2}\left[\omega_c^2+\omega_0^2\left(\sqrt a\pm \sqrt b\right)^2\right]^{1/2}$, $\ell=\sqrt{\frac{\hbar}{m\Omega}}$, $\Omega=\sqrt{\omega_0^2+\frac{1}{4}\omega_c^2}$ and $\omega_c=\frac{eB}{m}$ is the cyclotron frequency.

We now turn to the calculation of the phonon induced spin relaxation rate in between two lowest energy  states in QDs. The interaction between electron and piezo-phonon can be written as~\cite{khaetskii00,khaetskii01}
\begin{equation}
u^{\mathbf{q}\alpha}_{ph}\left(\mathbf{r},t\right)=\sqrt{\frac{\hbar}{2\rho V \omega_{\mathbf{q}\alpha}}} e^{i\left(\mathbf{q\cdot r} -\omega_{q\alpha} t\right)e A_{\mathbf{q}\alpha}b^{\dag}_{\mathbf{q}\alpha}} + H.c.
\label{u}
\end{equation}
Here, $\rho$ is the crystal mass density, $V$ is the volume of the QDs, $b^{\dag}_{\mathbf{q}\alpha}$ creates an acoustic phonon with wave vector $\mathbf{q}$ and polarization $\hat{e}_\alpha$, where $\alpha=l,t_1,t_2$ are chosen as one longitudinal  and two transverse modes of the induced phonon  in the dots. Also,  $A_{\mathbf{q}\alpha}=\hat{q}_i\hat{q}_k e\beta_{ijk} e^j_{\mathbf{q}\alpha}$ is the amplitude of the electric field created by phonon strain, where $\hat{\mathbf{q}}=\mathbf{q}/q$ and $e\beta_{ijk}=eh_{14}$ for $i\neq k, i\neq j, j\neq k$. The polarization directions of the induced phonon are $\hat{e}_l=\left(\sin\theta \cos\phi, \sin\theta \sin\phi, \cos\theta \right)$, $\hat{e}_{t_1}=\left(\cos\theta \cos\phi, \cos\theta \sin\phi, -\sin\theta \right)$ and $\hat{e}_{t_2}=\left(-\sin\phi, \cos\phi, 0 \right)$. Based on the Fermi Golden Rule, the phonon induced spin transition rate in the QDs is given by~\cite{sousa03,khaetskii01}
\begin{equation}
\frac{1}{T_1}=\frac{2\pi}{\hbar}\int \frac{d^3\mathbf{q}}{\left(2\pi\right)^3}\sum_{\alpha=l,t}\arrowvert M\left(\mathbf{q}\alpha\right)\arrowvert^2\delta\left(\hbar s_\alpha \mathbf{q}-\Delta\right),
\label{1-T1}
\end{equation}
where  $s_l$,$s_t$ are the longitudinal and transverse acoustic phonon velocities in QDs. The matrix element $M\left(\mathbf{q}\alpha\right)$ for the spin-flip rate between the Zeeman sublevels with the emission of phonon $\mathbf{q}\alpha$ has been calculated perturbatively.~\cite{khaetskii01,stano06}

\begin{figure}
\includegraphics[width=9cm,height=7cm]{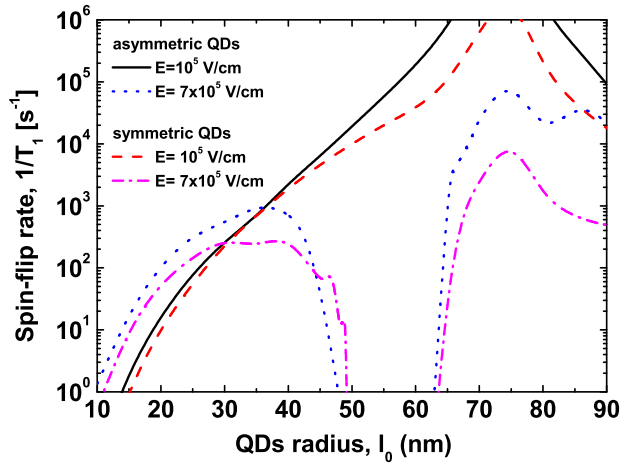}
\caption{\label{fig2}
(Color online) Phonon induced spin-flip rate due to spin-orbit admixture mechanism as a function of QDs radius in asymmetric QDs (solid and dotted lines). As a reference, we also plotted spin-flip rate vs. QDs radius for symmetric QDs (dashed and dashed dotted lines). We choose the potentials characterized by $a=0.5$ $\&$ $b=2$ for asymmetric QDs and a=b=1 for symmetric QDs. Also we choose $B=1$ T. As we see, spin-flip rate increases approximately by one half order of magnitude in asymmetric QDs.
}
\end{figure}
\begin{figure}
\includegraphics[width=9cm,height=5.5cm]{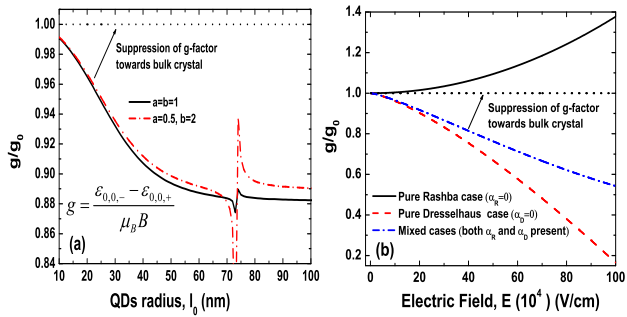}
\caption{\label{fig3}
(Color online) (a) The anisotropic effect on the g-factor vs. QDs radius at the potentials characterized by $a=b=1$ (solid line) for isotropic QDs and $a=0.5,b=2$ (dashed-dotted line) for anisotropic QDs. We choose $E=10^5$ V/cm and $B=1$T. Anisotropic potential gives the  suppression of  the g-factor towards bulk crystal and hence reduces the level crossing point to lower QDs radius. Accidential degeneracy appears in the range of $70-80$ nm QDs radius which gives the cusp like structure in the  spin-flip rate (see Refs.~\onlinecite{bulaev05,bulaev05a,stano05}). (b) The interplay between Rashba and Dresselhaus spin-orbit couplings on the g-factor vs. the electric field in QDs induces the anisotropic effect due to the suppression of the g-factor towards bulk crystal. Here, we choose $\ell_0=20$ nm, $B=1$ T and $a=b=1$.
}
\end{figure}

In Fig.~\ref{fig1}, we quantify the influence of the Rashba-Dresselhaus spin-orbit admixture mechanism on the phonon induced spin-flip rate as a function of quantum dot radius for symmetric QDs. For the electric fields $E=10^4$ and $10^5$ V/cm (shown by solid and dashed lines), we find that the transition time between spin up and down states increases with the  increase in QDs radius and gives the level crossing point at $\ell_0=73$ nm. For the case $E=7\times 10^5$ V/cm (shown by dashed dotted line), the spin relaxation rate in QDs starts decreasing at $\ell_0=35$ nm because Zeeman spin splitting energy is very small which implies small phonon density of states. It becomes  negligible at $\ell_0=50$ nm, and thus the spin relaxation rate turns  to be zero. The spin relaxation time starts increasing at $\ell_0=65$ nm. However, in this regime, the opposite spin state is dominating because the g-factor of an electron spin states changes their sign (see inset plots). This is an important result for the design of   spin based logic  devices. Indeed, in GaAs/AlGaAs QDs, wavefunctions of electrons penetrate from GaAs QDs to AlGaAs barrier with the application of gate controlled electric fields where the g-factor of an electron changes its sign.~\cite{jiang01,yang05,chang03}
The level crossing takes into account, due to mixing, the Zeeman spin states $|0,0,->$ and $|0,1,+>$ in QDs.
The crossing point is theoretically investigated  by the condition~\cite{bulaev05,bulaev05a} $\varepsilon_{0,0,-}=\varepsilon_{0,1,+}$ i.e., $\hbar\left(\omega_+-\omega_-\right)=|g_0|\mu_B B$ (see Eq.~\ref{epsilon-a}). Substituting $B=1$ T in the above condition, gives the crossing point at $\ell_0=73$ nm. Theoretically investigated level crossing point is in agreement with the numerically investigated values in the spin flip rate (see Fig.~\ref{fig1}). It can be seen that enhancement in the spin-flip rate occurs with the increase in electric fields. The level crossing point in the spin-flip rate is not affected by the electric fields which tells us that the level crossing  point found in the spin flip rate  is a purely orbital effect and is independent of the Rashba-Dresselhaus spin-orbit interaction.

Fig.~\ref{fig2} explores the influence of  anisotropic effects on the spin-flip rate vs. QD radius for the electric fields $E=10^5$ and $E=7\times 10^5$ V/cm. It can be seen that the anisotropic potential ($a=0.5,b=2$) enhances the spin-flip rate by approximately one half order of magnitude compared to that of symmetric potentials  ($a=b=1$). Note that we chose the above confining potentials in such a way that the area of the symmetric and asymmetric QDs are held constant.  The level crossing point determined by the condition $\hbar\left(\omega_+-\omega_-\right)=|g_0|\mu_B B$ for anisotropic QDs is smaller than for the case of isotropic QDs if we held the area of the QDs constant. The crossing point for symmetric QDs was first studied by the authors of Refs.~\onlinecite{bulaev05,bulaev05a}. However, in this paper,  we present the condition of the level crossing point for asymmetric QDs and by  utilizing both theoretical and  numerical  methods, we report  that  the anisotropic potential reduces the level crossing point to a smaller QD radius as well as to smaller magnetic fields. Similar to Fig.~\ref{fig1}, at the electric field $E=7\times 10^5$ V/cm for anisotropic QDs in Fig.~\ref{fig2} (dotted lines), small Zeeman energy implies negligible phonon density of states which gives zero spin flip rate at $\ell_0=46$ nm and the g-factor with opposite sign (spin states change their sign) is observed at $\ell_0=63$ nm. Note that these numerically estimated values (zero spin-flip rate and the g-factor with opposite sign) occur at smaller QD radii for anisotropic QDs than in isotropic QDs. This tells us  that the anisotropic potential leads to the quenching effect in the orbital angular momentum~\cite{pryor06} that pushes the  g-factor of an electron towards the bulk  crystal  which causes the level crossing point to occur at a smaller QD radius for anisotropic QDs (see Fig.~\ref{fig3}(a)).
From Fig.~\ref{fig3}(b), we see that the g-factor can be tuned with the spin-orbit interactions. Recall that the electric field controls the strength of Rashba and Dresselhaus spin-orbit couplings (see Eq.~\ref{coefficient-R-D}).
Here, we again see that the suppression of g-factor towards bulk crystal  induces  the anisotropic effect
due to the interplay between Rashba-Dresselhaus spin-orbit couplings.~\cite{liu11,sheng06}

To conclude, we have shown that  the electron spin states in the phonon induced spin-flip rate can be manipulated with the application of externally applied anisotropic gate potentials in QDs. The anisotropic potential causes the suppression of the g-factor towards bulk crystal that causes the enhancement of the spin-flip rate and reduces the level crossing point to a smaller QD radius. At sufficiently large electric fields, the phonon induced spin-flip rate can be tuned with Dresselhaus spin-orbit coupling by controlling the electron spin  states in a regime where the g-factor changes its sign.

This work has been supported by NSERC and CRC programs (Canada) and by
MICINN Grants No. FIS2008-
04921-C02-01 and FIS2011-28838-C02-01 (Spain).


%

\end{document}